\documentclass{aa}   
\usepackage{url}
\usepackage[colorlinks,citecolor=blue]{hyperref}\usepackage{graphicx}
\usepackage{natbib}
\bibpunct{(}{)}{;}{a}{}{,} 
\usepackage{txfonts}
\usepackage{graphicx}
\begin{document} 

   \title{Formation of giant radio sources in galaxy clusters}

   \subtitle{}

  \author{ Xiaodong Duan\inst{1,2}, Linhui Wu\inst{3}, Ruiyu Zhang\inst{1,2}, Jiawen Li\inst{4}  } 
  
    \institute{School of Physics, Henan Normal University, Xinxiang 453007, People’s Republic of China; \\
    \email{duanxiaodong@htu.edu.cn, zhangruiyu@htu.edu.cn}   
    \and {Center for Theoretical Physics, Henan Normal University, Xinxiang 453007, People’s Republic of China; }
      \and
      Shanghai Astronomical Observatory, Chinese Academy of Sciences, 80 Nandan Road, Shanghai, 200030, \\ People’s Republic of China; \email{wulinhui@shao.ac.cn}
     \and
      Department of Astronomy, School of Physics and Astronomy, Key Laboratory of Astroparticle Physics of Yunnan Province, \\Yunnan University, Kunming 650091,
People’s Republic of China;  \email{jwliynu@ynu.edu.cn}
       }

\titlerunning{Formation of Giant Radio Sources in Galaxy Clusters}
\authorrunning{Duan et al.}
     
  \abstract
   {The number of observed giant radio sources (GRSs) has increased significantly in recent years, yet their formation mechanisms remain elusive. The discovery of giant radio galaxies within galaxy clusters has further intensified the ongoing debates. }
   {We focus on the impact of jet properties, including jet power, energy components, and magnetic field structure, on the formation of GRSs within galaxy clusters. }
   {We utilize magnetohydrodynamic simulations to investigate the formation of GRSs in cluster environments. To avoid confusing the effects of power and total energy injection, we hold the energy of jet outbursts fixed and study the effect of power by varying the active duration of the jets. Furthermore, we examine the roles of magnetic, thermal, and kinetic energy components by adjusting their fractions in the jets. Additionally, we calculate radio emission for comparison with observations in the radio power-linear size diagram (P-D diagram). Finally, we also study the energy transport processes of different jets.}
   {We find the "lower power-larger bubble" effect: when the total jet energy is fixed, lower power jets tend to produce larger radio sources. Regarding different energy components, jets dominated by toroidal magnetic field energy generate larger radio sources than kinetic and thermal energy-dominated jets. Conversely, strong poloidal magnetic fields hinder radio lobe growth. When injecting $2.06 \times 10^{59}$ erg into a $10^{14}$ solar mass halo, only jets with powers of approximately $10^{-4}$ to $10^{-3}$ Eddington luminosity efficiently traverse the observational region in the P-D diagram.}
   {Our findings suggest that energetic, long-lasting (low-power), continuous jets endowed with significant toroidal magnetic fields facilitate the formation of GRSs in cluster environments. However, although the jets with significantly lower power can generate substantially larger radio sources, their faintness may render them unobservable. } 

   \keywords{Galaxies: active -- Galaxies: clusters: general -- Galaxies: jets -- Radio continuum: galaxies -- Magnetohydrodynamics
               }

   \maketitle
%

\section{Introduction}

In recent years, the number of observed giant radio sources (GRSs, defined to be >0.7 Mpc in size) has soared due to the low frequency radio surveys such as the LOFAR Two Metre Sky Survey (LoTSS) \citep{dabhade23, oei23, mostert24}. The largest GRS discovered to date spans approximately 7 Mpc \citep{oei24b}. The formation of GRSs has been debated for a long time, with several possibilities suggested in the literature: (1) GRSs are formed in low-density environments; (2) GRSs are formed due to high jet power; (3) recurrent jets form the GRSs \citep{dabhade23}. The idea of a low-density region as the primary mechanism is popular, as most GRSs appear to be found in such environments \citep{dabhade23}. However, recent observations have found that GRSs do not preferentially reside in low-density environments within their respective samples \citep{lan21, sankhyayan24, oei24a}. Furthermore, there are also GRSs identified within galaxy clusters \citep{dabhade20b, dabhade20a, tang20}. Consequently, the environmental dependence of GRSs remains uncertain, particularly regarding their formation within galaxy clusters.

On the other hand, numerical simulations indicate that the morphologies and scales of active galactic nucleus (AGN) bubbles (or radio lobes) are significantly influenced by jet properties \citep{guo15, guo20, english19, duan20, chen23}. In hydrodynamic simulations, very high-power jets can trigger strong outer shock dissipation, making it difficult to form large radio bubbles. Furthermore, thermalized strong jet outbursts may even result in bubbles that reside exclusively at the center of clusters \citep{duan20}. These results motivate us to investigate whether the formation of GRSs is contingent upon jet properties.

The jet parameters can be investigated through various methods in simulations, for instance, by altering jet properties while keeping jet power constant \citep{yates18} or by maintaining a specified maximum lobe length \citep{english19}. Unlike these previous studies, we keep the total energy of jet outbursts constant and study the effect of power by varying the active duration of the jets. This approach allows us to avoid confusing the effects of power and total energy injection. We also examine the impact of different energy components by controlling the proportions of magnetic energy, thermal energy, and kinetic energy injected through the jets.

The remainder of this paper is structured as follows. In Sect. \ref{sec2}, we outline the basic equations and our methodology. In Sects. \ref{sec3.1} and \ref{sec3.2}, we investigate the dependence of the scale of the radio sources on the parameters of the jets. Furthermore, in Sect. \ref{sec3.3}, we compare the radio sources from our simulations with observations in the P-D diagram (radio power-linear size diagram; \cite{baldwin82}). Subsequently, in Sect. \ref{sec4}, we delve into some details, particularly the energy transport processes of different jets. Lastly, in Sect. \ref{sec5}, we summarize our key findings.  

\section{Methods}
\label{sec2}
\begin{table}
\caption{List of Our Jet Parameters} 
\label{table1} 
\begin{tabular}{l c c c c c c} 
\hline\hline 
Run  & $\rm t_{jet}/Myr$ & $\rm P_{jet}/P_{fb}$ & $\rm f_{m}$   &   $\rm f_{th} $   &   $\rm \alpha_{p}$  \\ 
\hline 
t5                 &   5       &  10   & 0.05   &  0.05   & 0.3  \\ 
t5fth8           &   ---     &   ---    &0.05   &  0.8     & ---    \\
t5fm8           &   ---     &   ---    & 0.8     &  0.05   & ---     \\
t5fm8bp3     &   ---     &   ---    & 0.8     &  0.05   & 3     \\
t25               &   25    &    2    &0.05   &  0.05   & 0.3 \\ 
t25fth8         &   ---     &   ---    & 0.05    &  0.8     & ---     \\
t25fm8         &   ---     &   ---    & 0.8      &  0.05   & ---     \\
t25fm8bp3   &   ---     &   ---    & 0.8      &  0.05   & 3     \\
t50                &   50    &  1      & 0.05   &  0.05   & 0.3 \\ 
t50fth8          &   ---    &   ---    & 0.05   &  0.8     & ---     \\
t50fm8          &   ---    &   ---    & 0.8     &  0.05   & ---     \\
t50fm8bp3    &   ---    &   ---    & 0.8     &  0.05   & 3     \\
t200             &   200   & 0.25   & 0.05   &  0.05   & 0.3 \\ 
t200fth8       &   ---     &   ---    &0.05   &  0.8     & ---     \\
t200fm8       &   ---     &   ---    & 0.8     &  0.05   & ---     \\
t200fm8bp3  &   ---    &   ---    & 0.8     &  0.05    & 3     \\
t1200fm8     & 1200  &  0.04    & 0.8     &  0.05    & 0.3     \\
\hline 
\end{tabular}
\tablefoot{Jet parameters for our simulations. The parameters encompass the jet duration $t_{\rm jet}$, the jet power $P_{\rm jet}$ normalized by $P_{\rm fb}$, the magnetic energy fraction at the jet base $f_{\rm m}$, the thermal energy fraction at the jet base $f_{\rm th}$, and the parameter $\alpha_p$ of the magnetic field structure, which determines the ratio between the poloidal and toroidal magnetic fields. The symbol “—” indicates that the corresponding parameter shares the same value as the one immediately above it.}
\end{table}

\subsection{ Basic equations and numerical setup}

We solve the ideal magnetohydrodynamic (MHD) equations numerically in a 2.5-dimensional cylindrical coordinate system using the code MPI-AMRVAC 2.0 \citep{xia18},
\begin{eqnarray}  
\frac{\partial \rho}{\partial t} + \nabla \cdot( \rho \textbf{v}) = \dot{\rho}_{\text{inj}},
\end{eqnarray}
\begin{eqnarray}  
\frac{\partial (\rho \textbf{v}) }{\partial t} + \nabla \cdot ( \rho \textbf{vv}  - \textbf{B} \textbf{B} ) +\nabla (p + \frac{B^2}{2}) + \rho \nabla \Phi= \dot{\rho}_{\text{inj}} \textbf{v}_{\text{inj} },
\end{eqnarray}
\begin{eqnarray}  
\frac{\partial e }{\partial t} + \nabla \cdot [ (e + p + \frac{B^2}{2}) \textbf{v} - \textbf{B} \textbf{B} \cdot \textbf{v} ] + \rho \textbf{v} \cdot \nabla \Phi= \dot{e}_{\text{inj}},
\label{eq-e} 
\end{eqnarray}
\begin{eqnarray}  
\frac{\partial \textbf{B} }{\partial t} + \nabla \cdot  (\textbf{v}  \textbf{B} - \textbf{B} \textbf{v} )  = \dot{\textbf{B} }_{\text{inj}}.
\label{eq-B} 
\end{eqnarray}
In the equations above, $\rho$, $\textbf{v}$, and $p$ represent the density, velocity, and thermal pressure, respectively. $e$ denotes the total energy density, which comprises thermal energy, kinetic energy, and magnetic energy. $\textbf{B}$ is the magnetic field in code units and can be converted to Gauss units via the relation $\textbf{B} (\text{G}) = \sqrt{4\pi} \textbf{B}$. $\Phi$ represents the gravitational potential. The terms on the right-hand sides signify the jet injection when the jet is active. These equations can be closed by the equation of state, 
\begin{eqnarray}  
p = (\gamma - 1) ( e -  \frac{\rho v^2}{2} - \frac{\textbf{B}^2} {2}),
\end{eqnarray}
where $\gamma $ =5/3 in this work. 

We simulate only the evolution of one side of the bipolar jets axisymmetrically. The computational domain is initially refined at different resolutions: 0.25 kpc from the origin to 200 kpc, 0.5 kpc from 200 kpc to 300 kpc, 1 kpc from 300 kpc to 400 kpc, and 2 kpc from 400 kpc to 800 kpc. We apply reflective boundary conditions at the inner boundaries and outflow boundary conditions at the outer boundaries. In this study, we adopt the TVDLF scheme \citep{keppens12}, which is combined with the 'mcbeta' slope limiter and second-order temporal discretization. We employ the 'linde' method to maintain $\nabla \cdot \textbf{B} = 0$. Further details regarding these methodologies can be found on the MPI-AMRVAC website.

\subsection{Gravitational potential and initial gas distribution}

We adopt a fixed gravitational potential contributed by three components: the dark matter halo with a mass of $10^{14}M_{\odot}$, the central galaxy with a mass of $2.84\times 10^{11}M_{\odot}$ and the central black hole with a mass of $5.16 \times 10^{8} M_{\odot}$. Furthermore, we establish the initial density and pressure distribution of the ambient gas by assuming hydrostatic equilibrium. The specifics of our gravitational potential model and initial gas configuration are detailed in Appendix \ref{app_galaxy}. 

\begin{figure*}[t]
\centering
\includegraphics[height=0.25\textheight]{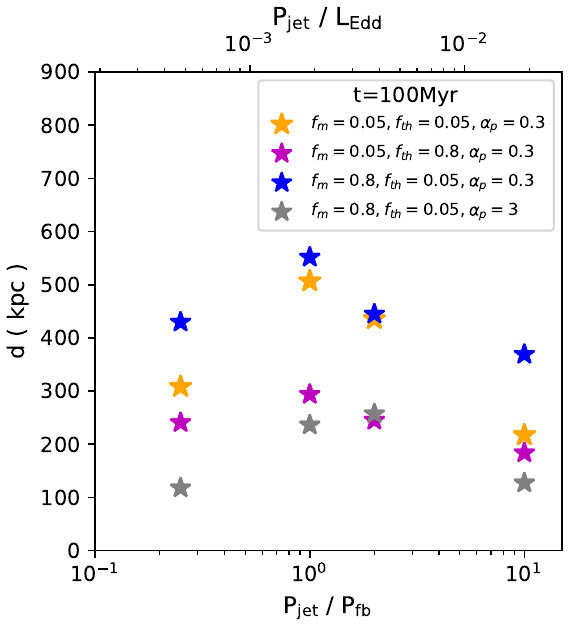}
\includegraphics[height=0.25\textheight]{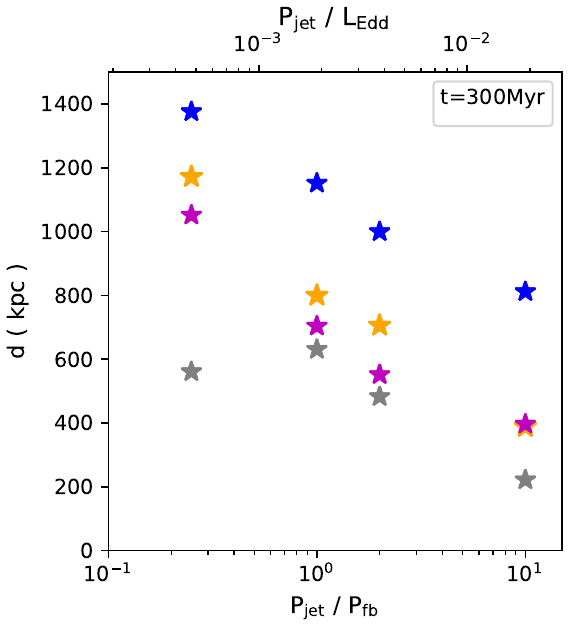}
\includegraphics[height=0.25\textheight]{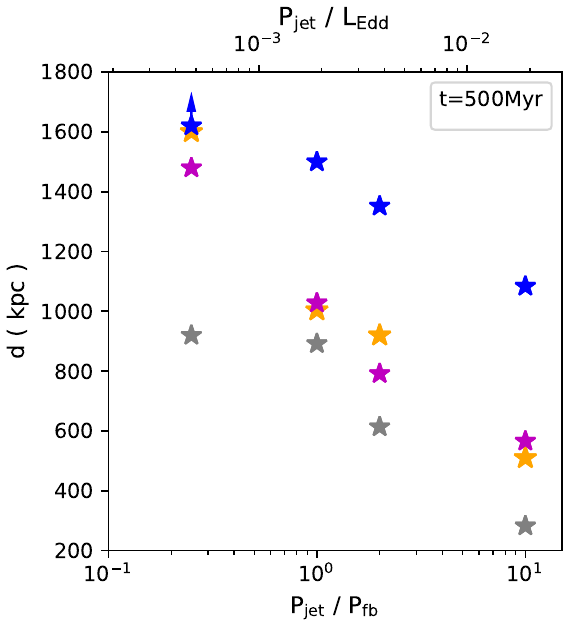}
\caption{Scales of the radio sources $d$ at different times (100Myr, 300Myr and 500Myr) varying with the jet powers. We determine the scales of the radio sources by doubling the distance from the heads of the one-side radio lobes to the center of galaxy. The stars are color-coded to represent distinct fractions of energy components and the magnetic structure parameter $\alpha_p$, which are indicated in the plot. The arrow indicates that the scale of the source is beyond the simulated region.}
\label{figdPt}
\end{figure*}

\subsection{Jet injection}
\label{subsec_jetinject}
In our simulations, AGN jets are activated from the start. During the active phase (when $t \leq t_{\text{jet}}$), a constant jet is injected along the $+z$ direction, injecting mass, momentum, and non-magnetic energy into a cylindrical region with a cross-sectional radius of $R_{0} = 1$ kpc and extending from $z = 0$ to a height of $z=h_{\text{\rm jet}} = 1$ kpc along the $+z$ axis. The method for injecting magnetic field is described in Appendix \ref{app_mag}. The jet velocity is set at a constant value of $v_{\text{jet}} = 0.1c$, where $c$ is the speed of light. The total energy injected into our simulated domain, representing the energy of one side of the jet, is $E_{\text{inj}} = 0.5 E_{\text{jet}}$, where $E_{\text{jet}}$ denotes the total energy of the bipolar jet system.

The energy of the jets can be estimated by assuming that the energy is extracted from the spin of the central black hole \citep{meier99, mcnamara09}:
\begin{eqnarray}  
E_{\text{jet}} \approx 1.6\times 10^{62} \left(\frac{ M_{\text{BH}} }{10^9 M_{\odot}}\right) a^2 \text{ erg} .
\end{eqnarray}
Observations have indicated that the typical value of the spin parameter $a$ for black holes in GRSs is approximately 0.05 \citep{dabhade20b}. In our simulations, the mass of the black hole is $M_{\text{BH}} = 5.16 \times 10^8 M_{\odot}$. Consequently, the energy that can potentially be extracted from this black hole is $E_{\text{jet}} \approx 2.06\times 10^{59} \text{ erg}$. Thus, the energy of the one-sided jet that we inject into our simulation is $E_{\text{inj}} = 1.03 \times 10^{59} \text{ erg}$, which represents half of the total bipolar jet energy.

The power of jets can be estimated by considering either feedback processes or accretion physics. When focusing on feedback processes, a typical power of jet outbursts is given by the equation:
\begin{eqnarray}  
P_{\text{fb}} = \frac{ E_{\text{jet}} }{ t_s },  
\end{eqnarray}
where $t_s$ represents the typical sound crossing time scale in the ambient gas, which is calculated within a characteristic feedback radius $R_{\text{fb}}$. $R_{\text{fb}}$ is the radius within which the initial thermal energy of the ambient gas is equal to the jet outburst energy \citep{duan20}. As $R_{\text{fb}}$  is determined by the initial gas distribution and jet injection energy, both of which are fixed in our simulations, it remains the same, at 31 kpc, for all our runs. In the context of this work, we have $t_s \approx 50 \text{ Myr}$ and $P_{\text{fb}} \approx 1.3 \times 10^{44} \text{ erg s}^{-1}$. The physical meaning of $P_{\text{fb}}$ lies in the fact that if the jet power, $P_{\text{jet}}$, significantly exceeds $P_{\text{fb}}$, an appreciable fraction of the power deposited by the jet will be dissipated in shocks. Furthermore, taking into account the Eddington luminosity of the black hole $L_{\text{Edd}} \approx 6.71 \times 10^{46} \text{ erg s}^{-1}$, the ratio of jet power to Eddington luminosity is $\lambda_{\text{jet}} = P_{\text{jet}} / L_{\text{Edd}}$. For the case where $P_{\text{jet}} = P_{\text{fb}}$, this ratio yields $\lambda_{\text{jet}} \approx 2 \times 10^{-3}$.

The power of AGN jets can also be related to the accretion rate as 
\begin{equation}  
P_{\text{jet}} = \eta_{\text{jet}} \dot{M} c^2.
\end{equation}
Considering the luminosity of the accretion flows, $L = \eta_{\gamma} \dot{M} c^2$, and the Eddington ratio, $\lambda_{\text{Edd}} = L/L_{\text{Edd}}$, the ratio of jet power to Eddington luminosity can be expressed as: 
\begin{equation}  
\lambda_{\text{jet}} = \lambda_{\text{Edd}} \frac{\eta_{\text{jet}}}{\eta_{\gamma}}.
\end{equation}
Here, the radiation efficiency $\eta_{\gamma} \approx 0.1$ and the jet efficiency $\eta_{jet} \lesssim 1.3 a^2 \approx 0.3\% $ for $a = 0.05$ \citep{Davis20}. Jets can be launched in both sub-Eddington hot accretion flows and super-Eddington accretion flows \citep{Tchekhovskoy15}.  For hot accretion flow $\lambda_{\rm Edd} \lesssim 2\%$ \citep{yuan14,yuan18},  we have $\lambda_{\rm jet} \lesssim 6.5 \times 10^{-4} $. On the other hand, for super-Eddington accretion flow with $\lambda_{\rm Edd} \sim 1$, we have $\lambda_{\rm jet} \lesssim 0.03$. To account for the uncertainties in the theory of jet launching, we will consider jet power within a broad range of approximately $10^{-5} \lesssim \lambda_{\text{\rm jet}} \lesssim 10^{-2}$ in this work.

We have conducted a series of numerical simulations with different jet parameters, focusing on situations where the total energy output remains fixed. The key parameters under investigation are the active duration $t_{\text{jet}}$, which governs the power of the jets, the thermal fraction $f_{\text{th}}$, the magnetic fraction $f_{\text{m}}$, and the parameter of magnetic field structure $\alpha_p$. We regulate the magnetic energy $E_{\text{m}}$ and thermal energy $E_{\text{th}}$ injected through the jets by defining the thermal fraction as $f_{\text{th}} = E_{\text{th}} / E_{\text{inj}}$ and the magnetic fraction as $f_{\text{m}} = E_{\text{m}} / E_{\text{inj}}$. Consequently, the fraction of kinetic energy is determined as $f_{\text{k}} = 1 - f_{\text{m}} - f_{\text{th}}$. To control the magnetic field energy and magnetic field structure injected into the jet base, we have developed a simple method for injecting the magnetic field, which is described in Appendix \ref{app_mag}. Our primary simulations encompass four sets with distinct jet powers, in addition to a low-power jet run (t1200fm8). The parameters for each run are detailed in Table \ref{table1}.

\section{Results}
\label{sec3}

\begin{figure*}[t]
\centering
\includegraphics[height=0.25\textheight]{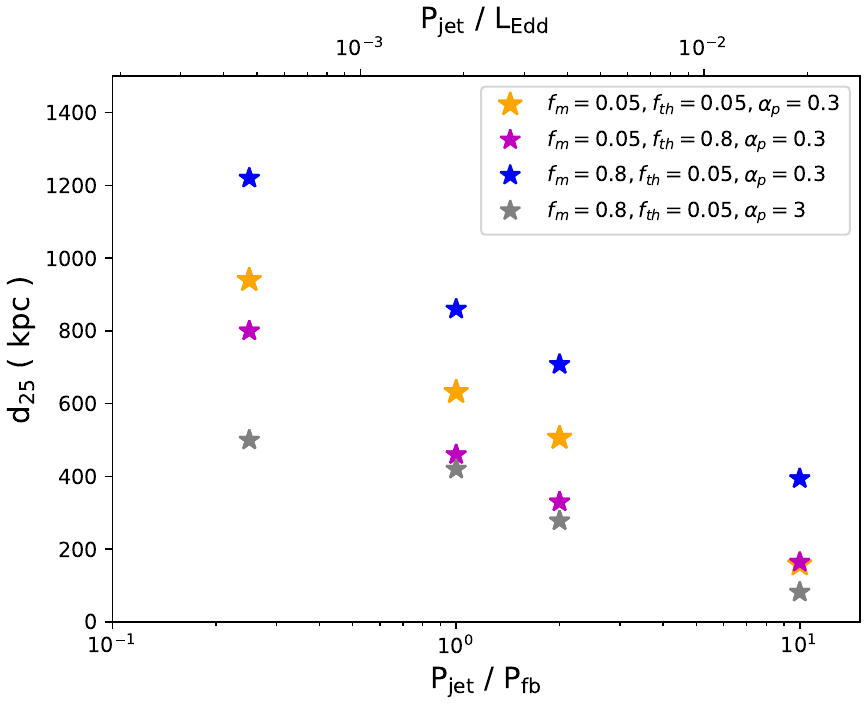}
\includegraphics[height=0.25\textheight]{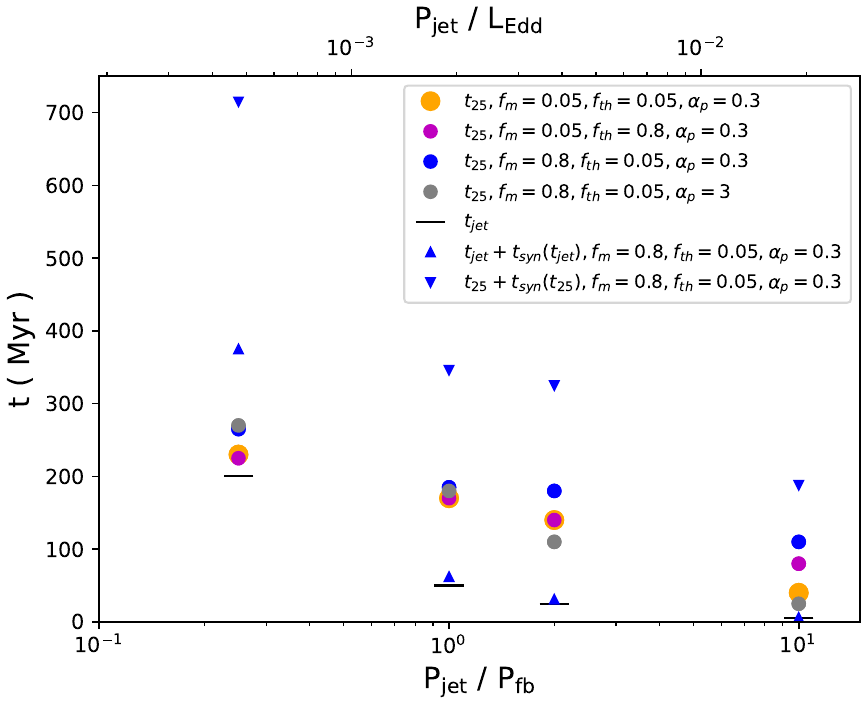}
\caption{Scales of the radio sources, $d_{25}$ at the radio power $10^{25} \rm{W Hz^{-1}}$ (left) and the corresponding times (right), varying with the jet powers, $P_{\rm jet}$, in simulations with diverse jet parameters, as shown in Table \ref{table1}. The dots in the right panel represent the time $t_{25}$ when the radio power of the source declines to $10^{25} \rm{W Hz^{-1}}$. The horizontal lines in the right panel mark the time $t_{\rm jet}$ when the corresponding jet stops being powered. The triangles denote $t_{\rm jet} + t_{\rm syn} (t_{\rm jet})$ (upward) and $t_{25} + t_{\rm syn}(t_{\rm 25})$ (downward), where $t_{\rm syn}(t)$ is the typical synchrotron cooling time  at 144(1+z)MHz for the volume averaged magnetic field strength corresponding to the time $t$, calculated using Eq. \ref{scooling}. The jet powers are normalized by the typical feedback power (bottom) and Eddington luminosity (top), as discussed in Sect. \ref{subsec_jetinject}. The stars are color-coded to represent distinct fractions of energy components and the magnetic structure parameter $\alpha_p$, as indicated in the plot. }
\label{figdP}
\end{figure*}

\subsection{"Lower power-larger bubble" effect}
\label{sec3.1}

The scales of the radio sources $d$ at different times (100 Myr, 300 Myr and 500 Myr) varying with the jet powers are shown in Fig. \ref{figdPt}. It is evident that at later stages (t=300Myr and 500Myr), the lower power jets generally yield larger radio sources (or jetted bubbles) compared to the high power jets. For the sake of brevity, this phenomenon will henceforth be referred to as the "lower power-larger bubble" effect in the remainder of this paper. The similar effect has been discovered in spherical outbursts \citep{duan20}. This phenomenon can be attributed to the fact that, initially, the total energy injected into the jet base by the lower power jets is substantially less than that injected by high power jets. However, when comparable amounts of energy have been injected, the "lower power-larger bubble" effect appears. Some details will further be discussed in Sect. \ref{sec4}. 

It is also interesting to see whether the "lower power-larger bubble" effect exists when considering the same radio power for different sources. In the left panel of Fig. \ref{figdP}, we present the scales of the radio sources varying with jet power, specifically at the instant when the sources have declined to radio power (luminosity) of $10^{25} \rm{W Hz^{-1}}$ at 144 (1+z) $\rm{MHz}$ for redshift z=0.3. The methodology for calculating radio emission is detailed in Appendix \ref{app_rad}. Notably, we can see a trend that the lower power jets result in larger radio sources. The corresponding time $t_{25}$ for the radio power of $10^{25} \rm{W Hz^{-1}}$ in different runs are shown in the right panel of Fig. \ref{figdP} (dots), and we can see they all occur after the jets are switched off ($t_{25}>t_{\rm jet}$ ). We also calculate the typical synchrotron cooling times for the toroidal magnetic field dominant runs at times $t=t_{\rm jet}$ and $t=t_{25}$. We can observe that, if the electrons cannot be adequately accelerated after the jets are switched off, they will quickly cool down, particularly for the high-power jets (upward triangles in the right panel of Fig. \ref{figdP}). This means that, the $t_{25}$ for the high-power jets ($t_{\rm jet}<200$ Myr or $P_{\rm jet}/P_{\rm fb}>0.25$) might be overestimated and when considering synchrotron cooling, the high-power jets will actually result in even smaller radio sources for the same radio power.      

\subsection{Behavior of magnetic field}
\label{sec3.2}

We also examine the impact of energy partition on the jet evolution. The results are presented in Fig. \ref{figdP}. Notably, jets dominated by toroidal magnetic energy ($f_{m} = 0.8$, $\alpha_p = 0.3$) yield larger radio sources compared to those dominated by thermal or kinetic energy ($f_{m} = 0.05$, $\alpha_p = 0.3$). This should be due to the magnetic tension force directed toward the jet axis (hoop stress) which can facilitate jet collimation, akin to mechanisms proposed to explain jet collimation near black holes \citep{Romero14}. This collimation results in slimmer lobes with reduced cross-sections and lower drag forces during propagation through the ambient gas. Furthermore, in Fig. \ref{fig_morph2} of Appendix \ref{app_rad}, we can see that a stronger magnetic field shields the lobes from Kelvin-Helmholtz instability and minimizes mixing with the ambient gas (comparing run t5, t5fth5 and t5fm8). Consequently, the magnetic field's significance can also extend to explaining the bent radio sources observed in galaxy clusters, which are influenced by intra-cluster weather \citep{gan17}. 
 
To study the effects of magnetic field structure in greater detail, we increase $\alpha_p$ from 0.3 to 3 in simulations where a strong magnetic field ($f_m = 0.8$) is injected. As shown in the runs with a stronger poloidal magnetic field ($\alpha_p = 3$ in Fig. \ref{figdP}), the growth of radio lobes is effectively suppressed (see also in \cite{chen23}). This phenomenon is likely attributed to the reduced magnetic tension force directed toward the jet axis and the enhanced magnetic tension force toward the galaxy's center within the lobe, resulting from the stronger poloidal magnetic field ($\alpha_p = 3$).

\subsection{Radio power of the simulated sources}
\label{sec3.3}

\begin{figure}[t]
\centering
\includegraphics[height=0.27\textheight]{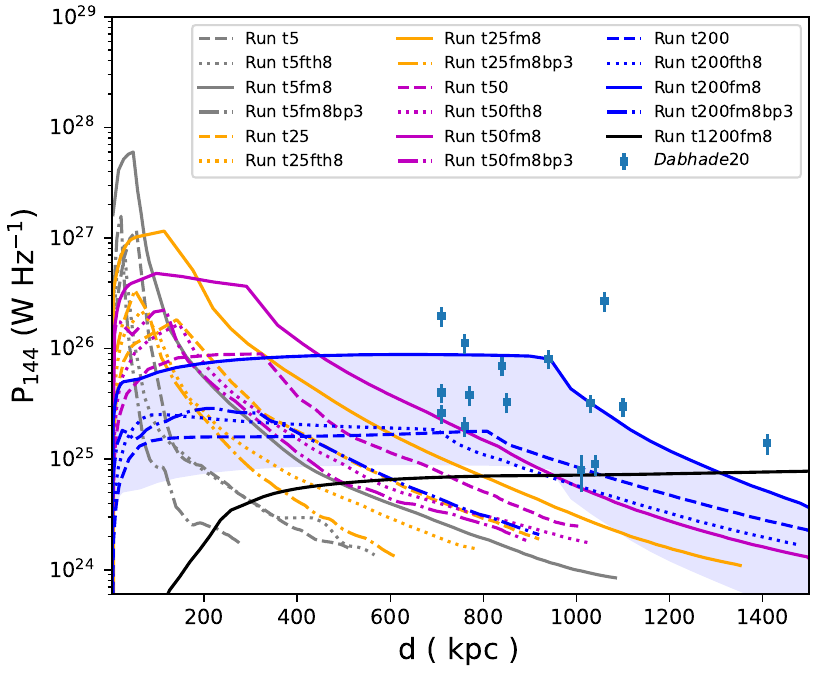}
\caption{Evolution of the simulated radio sources in the P-D diagram. Distinct colors represent different powers, while different line styles signify distinct fractions of energy components and the magnetic structure parameter, $\alpha_p$, as indicated in the plot. The blue shade covers regions with radio power one order of magnitude lower than that indicated by the blue solid line. The detailed methodology for calculating radio power is described in Appendix \ref{app_rad}. The data points utilized in this analysis are sourced from \cite{dabhade20a}. }
\label{figPD}
\end{figure}

Direct comparison between models and observations is challenging due to factors such as uncertainties of the properties of the non-thermal particles in lobes, the projection effect, and the quality of observed data \citep{hardcastle18}. Nevertheless, as an approximation, we can gain some useful insights into the jet properties of GRSs by neglecting these details. The evolution of radio sources in the P-D diagram for all our simulations is presented in Fig. \ref{figPD}. Comparing Fig. \ref{figPD} and Fig. \ref{figdE}, we can see that the radio powers decline significantly after the jet is switched off, and those produced by higher power jets decline earlier. The data points in Fig. \ref{figPD} represent the results of GRSs identified in clusters within the halo mass range of $0.7-2.0\times10^{14} M_{\odot}$, as observed in \cite{dabhade20a}. It is clearly demonstrated that only the low-power jets ($t_{\rm jet} = 200 \ \text{Myr}$) can readily traverse the observational region in the P-D diagram. A typical morphology of the radio sources produced by the low-power jets (run t200fm8) is shown in Fig. \ref{fig_morph1}.

On the other hand, although jets with significantly lower power can generate much larger radio sources, they may be too faint to observe. As illustrated in Fig. \ref{figPD}, the evolution of the lowest-power jet (run t1200fm8, represented by the black solid line) barely reaches the lower boundary of the observational region. This result implies that the jet powers of these radio sources lie approximately within the range of $10^{-4} - 10^{-3} L_{\rm Edd}$ (the corresponding ratios of the jet powers to the Eddington luminosity are shown in Fig. \ref{figdP}). Furthermore, the jet active time of run t1200fm8 is approximately the typical synchrotron cooling time (Eq. \ref{scooling}) of the electrons in the radio lobes, thus the outer part of the lobes might be radio dark if new acceleration of electrons is not adequate. Considering the very low density and pressure of the extreme low-power jets, the mixing between ejecta and surrounding medium might even cause them fail to form low density bubbles or observable radio lobes.     

\section{Discussion}
\label{sec4}

\subsection{Energy transport processes}
\label{sec4.1}

We examine energy transport processes in different runs before discussing in detail the production of large radio sources. We now focus on two sets of runs: the runs with the same energy fractions ($f_{\rm m} = 0.8$, $f_{\rm th} = 0.05$, $\alpha_{\rm p} =0.3$) but different powers and the runs with the same power ($t_{\rm jet} = 50\ \rm Myr$) but different energy fractions or magnetic field structure. The upper panels in Fig. \ref{figdE} show the evolution of the distance that the jetted bubble traveled (half of the dimension of radio sources) in these two sets of runs. The evolution of the total energy residing in jetted bubbles is also shown in the lower panels of \ref{figdE}. As an approximation, the bubble region is identified by a density lower than $5\times 10^{-27} \rm{g\ cm^{-3}}$ and a toroidal magnetic field stronger than $10^{-9} \rm{G}$. The dotted vertical lines show when the fixed total energy has been injected and the jet is turned off in different runs. We can see the scales of the jetted bubbles generally shows positive correlation with the total energy left in the bubbles particularly for the jets with the same energy makeup but different active times (powers). Notably, the high-power jet (run t5fm8) expels energy more rapidly upon jet cessation, leaving less energy within the bubble, which impedes the growth of the radio sources. Actually the most high-power jets can transmit energy outward through more potent shocks compared to low-power jets, which rely on less potent processes such as sound waves and slow expansions \citep{tang17, bambic19, duan20}.        

\begin{figure*}
\centering
\includegraphics[height=0.26\textheight]{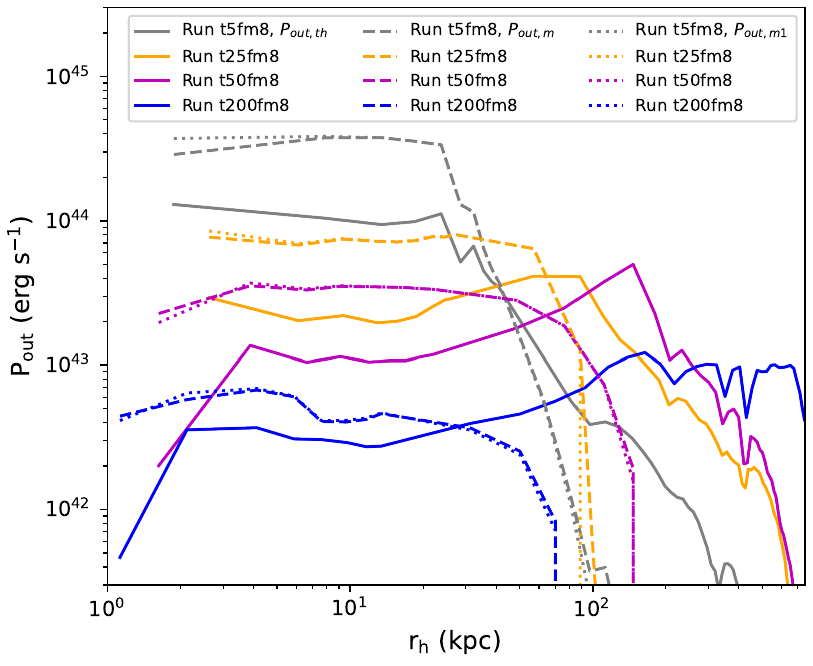}
\includegraphics[height=0.26\textheight]{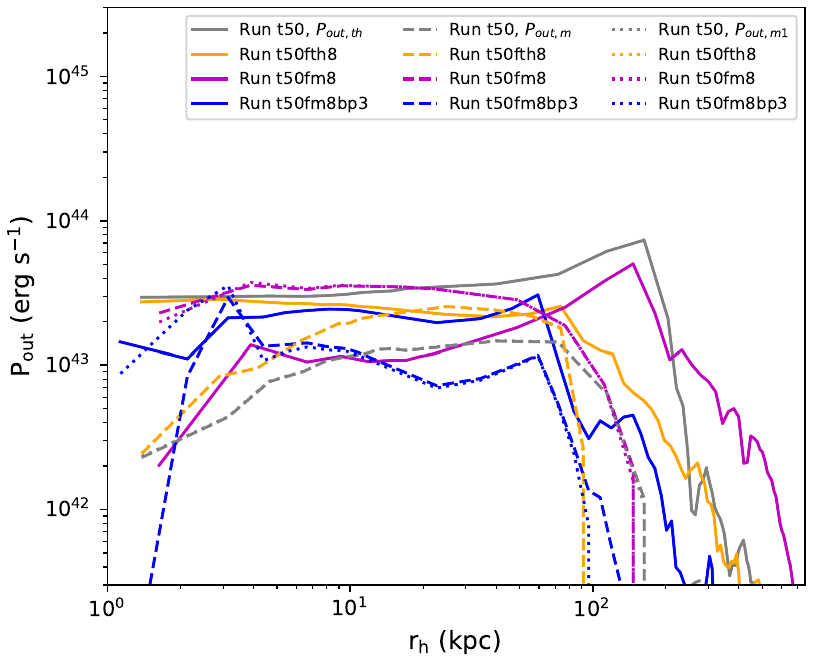}
\caption{ Thermal pressure power (solid lines), Maxwell stress power (dashed lines) and magnetic pressure power (dotted lines), calculated using Eqs. \ref{eq-pout-th}-\ref{eq-pout-m2}, varying with the traveling distance of lobes. Different colors represent the two sets of runs discussed in Sect. \ref{sec4.1}. }
\label{figEfluxi}
\end{figure*}

While neglecting the mixing processes, the energy outflow rate from the bubble should be equivalent to the power associated with  the work done by the bubble on the surrounding medium via thermal pressure and Maxwell stress (magnetic pressure plus magnetic tension), as shown in Eq. \ref{eq-pout2}. This power can be decomposed into three components: thermal pressure power $P_{out,th}$, magnetic pressure power $P_{out,m1}$, and magnetic tension power $P_{out,m2}$, which can be calculated as follows.   
\begin{equation}  
 P_{out,th} = \int_{S_{bub}} p\textbf{v} \cdot d\textbf{S}, 
\label{eq-pout-th}
\end{equation}  
\begin{equation}  
 P_{out,m1} = \int_{S_{bub}}  \frac{B^2}{2} \textbf{v} \cdot d\textbf{S}, 
\label{eq-pout-m1}
\end{equation}  
\begin{equation}  
 P_{out,m2} = -\int_{S_{bub}}  \textbf{v} \cdot \textbf{B} \textbf{B} \cdot d\textbf{S}. 
\label{eq-pout-m2}
\end{equation}  
The Maxwell stress power can be expressed as $P_{out,m} = P_{out,m1} + P_{out,m2}$. The evolution of thermal pressure power, Maxwell stress power and magnetic pressure power varying with the traveling distance of lobes is shown in Fig. \ref{figEfluxi}. The magnetic tension power can be determined by the difference between the Maxwell stress power and the magnetic pressure power, which can be negative for some cases (for example, during the early stage of run t50fm8bp3, as shown in the right panel of Fig. \ref{figEfluxi}). The thermal pressure power and Maxwell stress power in the higher power jet runs are both higher than those in lower power jet runs at the inner regions of clusters, as shown in the left panel. This means that higher power jets experience greater adiabatic losses in the inner regions of clusters, leaving less energy residing in the bubbles at later times, as discussed in the last paragraph. This should be related to the "lower power-larger bubble" effect. On the other hand, we can see that lower power jets can transport more energy outside in the outer region of clusters, as indicated by the left panel of Fig. \ref{figEfluxi}. Different energy makeups of jets do not result in as prominent differences as different powers, as shown in the right panel of Fig. \ref{figEfluxi}.

\begin{figure}[t]
\centering
\includegraphics[height=0.35\textheight]{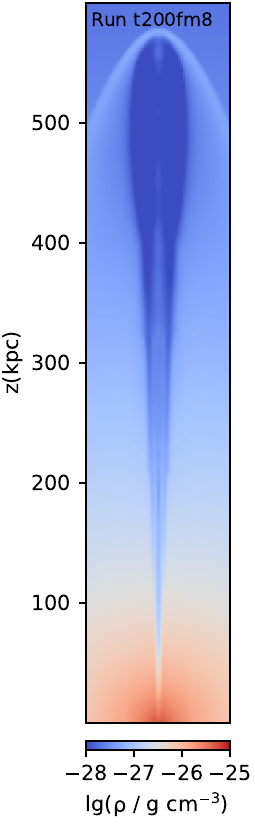}
\includegraphics[height=0.35\textheight]{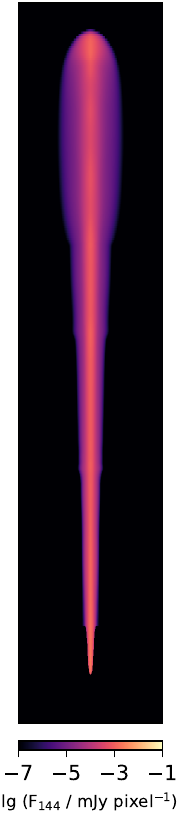}
\includegraphics[height=0.35\textheight]{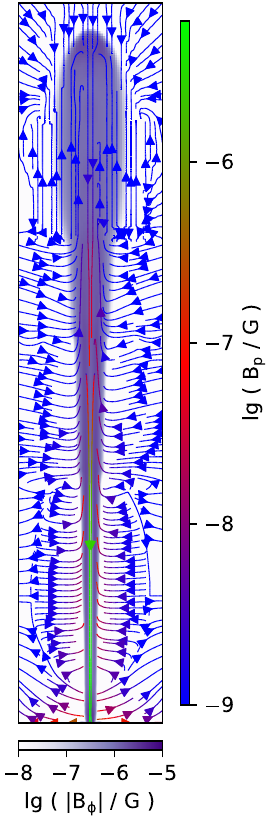}
\caption{Snapshots of density (left panel), radio flux (middle panel), and magnetic field (right panel) for run t200fm8 at $t = 250 \rm{Myr}$.   }
\label{fig_morph1}
\end{figure}

\subsection{Production of large radio sources }
\label{sec4.2}

\begin{figure*}[t]
\centering
\includegraphics[height=0.48\textheight]{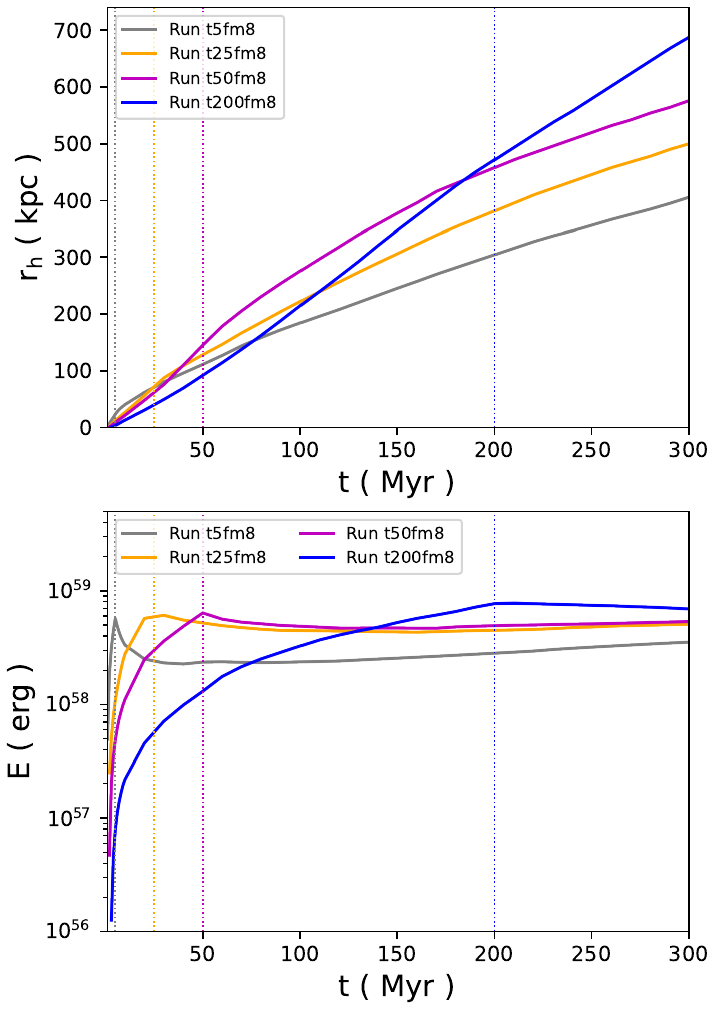}
\includegraphics[height=0.48\textheight]{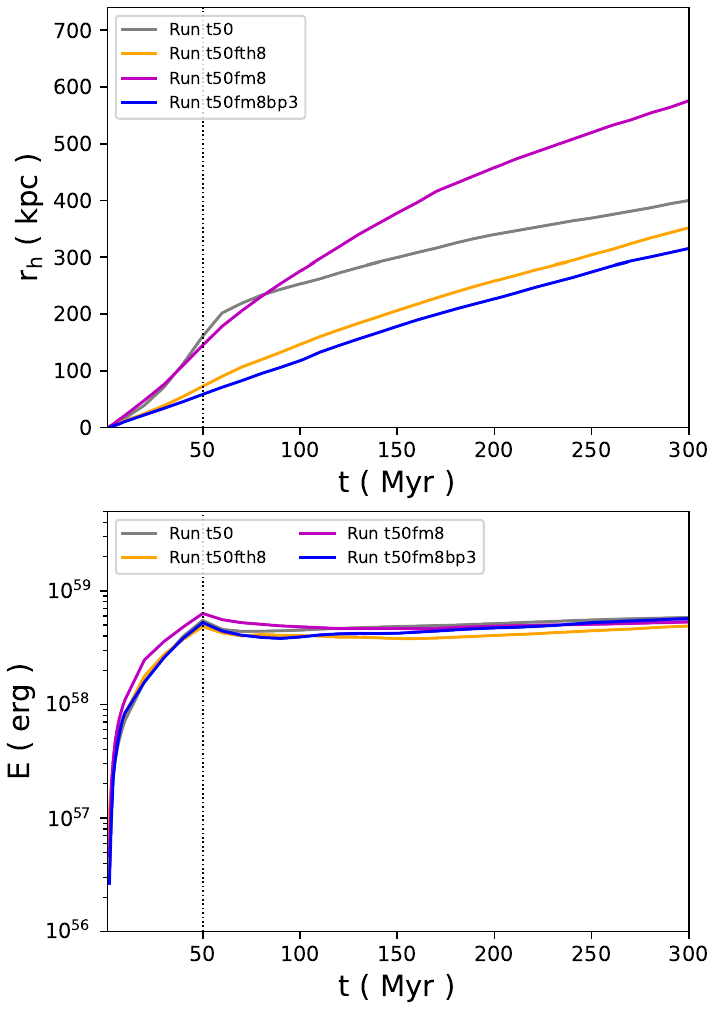}
\caption{Evolution of the traveling distance of the heads of bubbles (upper panels) and the energy contained within the jetted bubbles (lower panels) in the two sets of runs discussed in Sects. \ref{sec4.1} and \ref{sec4.2}. The bubble region is identified by a density lower than $5\times 10^{-27} \rm{g\ cm^{-3}}$ and a toroidal magnetic field stronger than $10^{-9} \rm{G}$. The subsequent mixing between the bubble region and the ambient gas may result in an upward trend in the curves of energy. Different colors of lines represent different runs as shown in the plots. The vertical lines denote when the jets are shut off.}
\label{figdE}
\end{figure*}

\begin{figure*}[t]
\centering
\includegraphics[height=0.48\textheight]{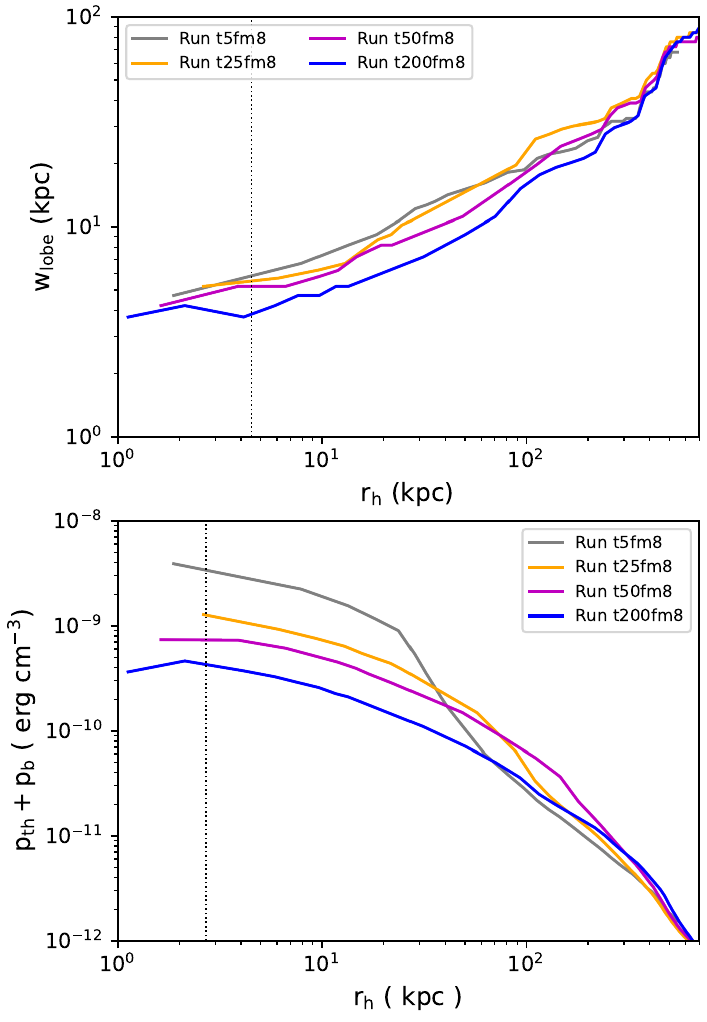}
\includegraphics[height=0.48\textheight]{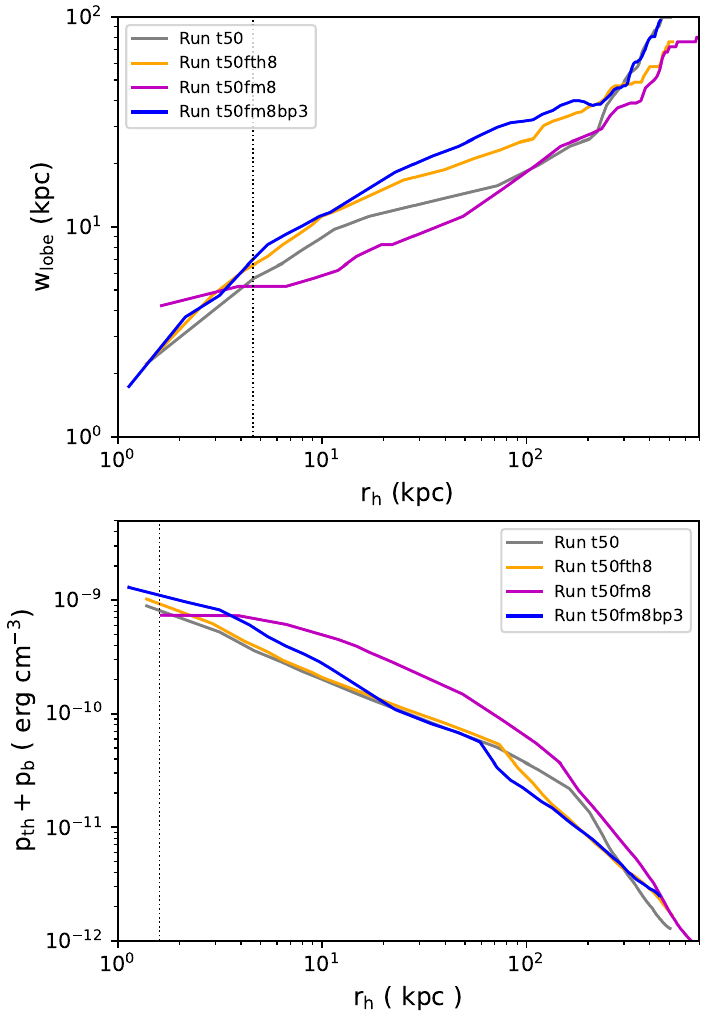}
\caption{Width (upper panels) and volume averaged internal pressure (including thermal and magnetic pressure, lower panels) of jetted bubbles varying with the traveling distance of the heads of bubbles in the two sets of runs discussed in Sects. \ref{sec4.1} and \ref{sec4.2}. Different colors of lines represent different runs, as shown in the plots. The vertical lines denote the points where anti-correlations between the width (or pressure) and the final scales of the radio sources are beginning to develop.}
\label{figwpthb}
\end{figure*}

To gain some insight into the production of large radio sources, we examine the evolution of the widths of the jetted bubbles and show the results in the upper panels of Fig. \ref{figwpthb}. We can see that, within the central region of the halos (for example,  r<100kpc), the widths of the bubbles generally exhibit an anti-correlation with the ultimate extent of the simulated radio sources (see Figs. \ref{figdPt} and \ref{figdE}). Furthermore, we check the volume-averaged internal pressure (magnetic pressure plus thermal pressure) within the bubbles and present the results in the lower panels of Fig. \ref{figwpthb}. We can see that internal pressures of bubbles show a positive correlation with the widths of bubbles in the innermost region of the halos. The dotted vertical lines show where the correlations have just begun to develop. This phenomenon is due to the fact that higher internal pressures within bubbles will cause faster transverse expansion, resulting in larger cross sections. The larger cross sections will make the jets much less penetrating and easier to be decelerated by the dense gas in the center of the halos. In the upper left panel of Fig. \ref{figwpthb}, we can observe that higher power jets generally produce wider bubbles at the center of the halos. Subsequently, there will be stronger interactions between the higher power jets and the surrounding gas, causing them to transfer more energy than lower power jets in the center of the halos, as discussed above. 

We then check the different results in run t50fm8 and t50fm8bp3 which are injected with the same energy fractions and power but have different magnetic structures. As the emerging jet is far from equilibrium with the surrounding gas, it will undergo rapid changes in its properties. Before the identified low-density bubbles are established, the jet in run t50fm8 should have experienced fiercer expansion than the jet in run t50fm8bp3, so that it is wider at the innermost region, as shown in the upper right panel of Fig. \ref{figwpthb}. This also causes the thermal and kinetic energy density in the jet in run t50fm8 to be less than that in run t50fm8bp3, as shown in the right panel of Fig. \ref{figei}. The rapid expansion process might be due to the more prominent hoop effect of the stronger toroidal magnetic field in run t50fm8, as it will make the ejecta interact strongly with the dense gas in front of the jet at the center of the halos. Then, when the channel is open, the jet in run t50fm8 becomes slimmer than the jet in run t50fm8bp3.

On the other hand, the inflated bubbles with stronger poloidal magnetic field and thermal energy (run t50fm8bp3 and t50fth8) also seem to experience stronger Kelvin-Helmholtz instability than the bubbles jetted by toroidal magnetic field dominant jets, as shown in Fig. \ref{fig_morph2}. This will make things more complicated, as the Kelvin-Helmholtz instability can trigger more prominent mixing with the surrounding medium. The mixing processes will actually make the low density bubbles lose more matter and energy than the stable ones, and finally reduce their potential to grow into large radio sources. This phenomenon is more pronounced in the case of higher power jets, as illustrated in Fig. \ref{fig_morph2}. We will address more details regarding the relationship between the structure of the magnetic field and the Kelvin-Helmholtz instability in future studies.   

 \subsection{Extended phenomena and limitations}
\label{sec4.3}
 
The large-scale magnetic field within black hole accretion flows is pivotal for jet launching and collimation \citep{cao11, chen21, lijiawen22}. Our simulations underscore the criticality of the toroidal magnetic field at kpc-scale in initiating and confining giant radio jets/lobes, as discussed in Sect. \ref{sec3.2}. Additionally, the wake flows of AGN bubbles are essential for elucidating metal-rich outflows and cold filaments in galaxy clusters \citep{duan18, duan24}. This study reveals that jetted bubbles can amplify the weak magnetic fields in the centers of clusters and ultimately result in stronger magnetic fields in their wake flows as shown in Fig. \ref{fig_morph1} and \ref{fig_morph2}, which may contribute to the seeding of magnetic fields in the circumgalactic and intergalactic media at later stages.            

The proportion of GRSs in galaxy clusters appears to be low \citep{dabhade20a}, suggesting their real-world formation may be challenging. In our simulations, with a fixed total injection energy, GRS formation is also contingent upon the total energy injection, and insufficient energy, even at the same power, will fail to yield GRSs. Our results suggest that a long-term stable jet output from the supermassive black hole's accretion system is necessary for GRS production. Additionally, GRSs generated by excessively low-power jets (run t1200fm8) may be too dim for detection, as discussed in Sect. \ref{sec3.3}.

The morphology of GRSs is diverse in observations \citep{dabhade20a}, and thus a detailed comparison with our simulations is beyond the scope of this work. Neither do we apply intra-cluster weather effects in our simulations, which explains the bent radio lobes typically seen in clusters \citep{gan17}. We observe that many giant radio sources extend along the jet axis \citep{dabhade20a}, a feature readily replicated in our simulations, particularly with strong toroidal magnetic field injections (Figs.s \ref{fig_morph1} and \ref{fig_morph2}). Another notable property is the significant decrease in radio flux at the edges in our simulations (middle panel, Fig. \ref{fig_morph1}), consistent with observations of some large radio lobes \citep{wu20}. Furthermore, the gas environment can influence GRS morphology. For instance, in gas halos with lower baryon fractions, jetted lobes may expand more readily, adopting fatter shapes. We aim to delve into the effects of different gas environments in future studies. The velocity will also influence the morphology of radio sources. If higher velocities (for example,  fully relativistic) of jets are considered, the density of the jets will be much lower for the same power or kinetic energy flux. This will render the jets less penetrating and produce flatter lobes, as demonstrated in \cite{english16}. On the other hand, full three-dimensional magnetohydrodynamic simulations of jets may encounter kink instability \citep{mignone10} and three-dimensional turbulence, both of which will pose difficulties for the growth of radio sources. The turbulence triggered by Kelvin-Helmholtz and Rayleigh-Taylor instabilities in two dimensions cannot experience the energy cascade to small scales as it does in three dimensions, which will result in different entrainment properties and morphologies of the jetted lobes \citep{massaglia16}.

\section{Conclusions}
\label{sec5}
We utilize magnetohydrodynamic simulations to study the formation of GRSs within galaxy clusters. To avoid confusing the effects of power and total energy injection, we hold the energy of jet outbursts fixed and study the effect of power by varying the active duration of the jets. We also examine the roles of magnetic, thermal, and kinetic energy components by adjusting their fractions in the jets. Additionally, we calculate radio emission for comparison with observations in the P-D diagram. The main results are summarized as follows:

(i) "Lower power-larger bubble" effect. For a fixed total energy in jet outbursts, we present the scaling of radio source sizes varying with jet power, particularly at the instant when the sources have declined to a common radio power of $10^{25} \rm{W Hz^{-1}}$ at 144(1+z) MHz. We have found a trend that the jets with lower power yield larger radio sources.

(ii) Behavior of magnetic field. When the energy is injected predominantly in the form of toroidal magnetic field, the jets tend to produce larger radio sources. If instead the poloidal field injected is stronger, the growth of the radio lobes is suppressed.

(iii) P-D diagram. When injecting jet outbursts of $E_{\text{jet}} = 2.06 \times 10^{59}$ erg into a halo of $10^{14} M_{\odot}$, only jets with powers approximately within the range of $10^{-4}$ to $10^{-3} L_{\rm Edd}$ and magnetic energy dominated by the toroidal component can readily traverse the observational region in the P-D diagram. Thus, our results suggest that energetic, long-term (low power), continuous jets with significant kpc-scale toroidal magnetic fields facilitate GRSs formation in cluster environments. On the other hand, although the jets with significantly lower power can generate much larger radio sources, they may be too faint to be observable.

(iv) Energy transport processes. Higher power jets generally produce jetted bubbles with higher internal pressure, which makes them wider and less penetrating through the centers of clusters, ultimately producing smaller radio sources. At the same time, higher power jets expel energy more rapidly at the centers of galaxy clusters through adiabatic losses, resulting from the work done on the surrounding medium by thermal pressure and Maxwell stress. This process leaves less energy within the bubbles produced by higher power jets, thereby impeding the growth of the radio sources.

We primarily focus on the impact of jet properties on the formation of GRSs within cluster environments. However, this work does not delve into other potential factors that may influence the formation processes and properties of GRSs, such as the effects of different gas environments, the extreme relativistic ejecta, the three-dimensional turbulence, and the kink instability during jet propagation. Our future studies aim to comprehensively explore these effects.

\begin{acknowledgements}
We thank the anonymous referee for the constructive reports, particularly for the suggestions regarding the addition of discussions. This work was supported by the High Performance Computing Center of Henan Normal University. R.Z. was supported by the China Postdoctoral Science Foundation (No. 2023M731014). J.L. was supported by the NSFC (12303020), the Yunnan Fundamental Research Projects (NO.202401CF070169), and the Xingdian Talent Support Plan - Youth Project.
\end{acknowledgements}

%
%
\bibliographystyle{aa} 
\bibliography{bibliography.bib} 
\nocite{*}


\begin{appendix}

\section{Gravitation and gas environment}
\label{app_galaxy}

\begin{figure}[!htb]
\centering
\includegraphics[height=0.45\textheight]{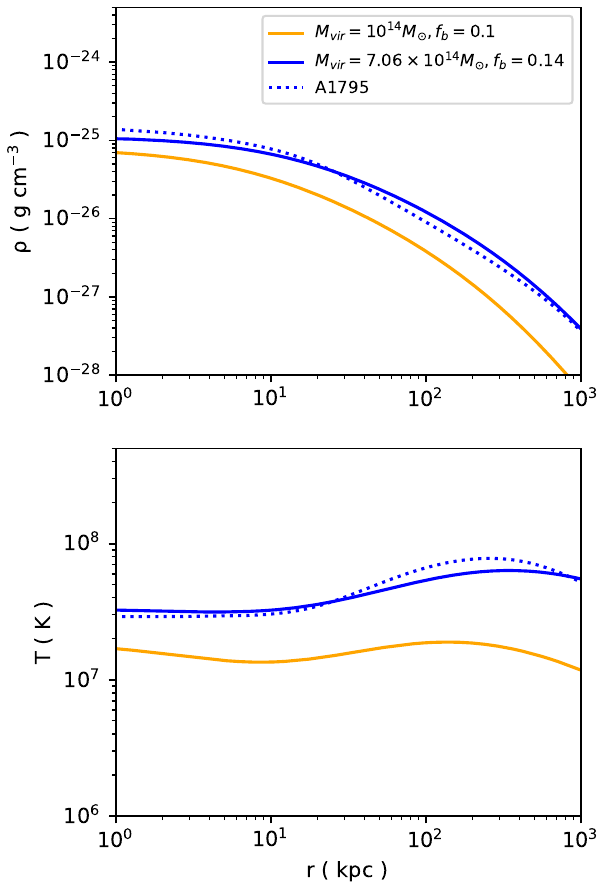}
\caption{ Profiles of density and temperature are presented for the virial mass of $M_{\text{vir}} = 10^{14} M_{\odot}$ (orange solid lines,  $f_b = 0.1$, adopted in this work) and $M_{\text{vir}} = 7.06 \times 10^{14} M_{\odot}$ (blue solid lines, $f_b = 0.14$), the latter being contrasted with the fitting results of galaxy cluster A1795 (blue dashed line, \citep{guo18}. } 
\label{figrhoT}
\end{figure}

We modify the models presented in \cite{guo18} and \cite{fang20} to establish the gravitational potential and gas environment for giant elliptical galaxies in galaxy clusters. For the dark matter gravitational potential, we adopt the Navarro–Frenk–White (NFW) profile \citep{navarro97}:

\begin{eqnarray}  
\Phi_{\text{dm}} = -\frac{2GM_0}{r_\text{s}} \frac{\ln (1+r/r_\text{s})}{r/r_\text{s}}
\end{eqnarray}
Here, $M_0$ and $r_\text{s}$ are determined by the virial mass of the dark matter halo $M_{\text{vir}}$, as follows:
\begin{eqnarray}  
M_0 = \frac{M_{\text{vir}}/2} { \ln(1+C) -C/(1+C) },
\end{eqnarray}
\begin{eqnarray}  
R_{vir} = 206 \left(\frac{M_{\text{vir}}}{10^{12}M_{\odot}}\right)^{1/3} \text{kpc},  
\end{eqnarray}
\begin{eqnarray}  
r_s = \frac{R_{\text{vir}}}{C} . 
\end{eqnarray}
In this study, we adopt a concentration parameter $C=4.06$ for a virial mass of $M_{\text{vir}} = 10^{14}M_{\odot}$ for the halo of clusters (typical value in \cite{dabhade20a}).

The potential of the central galaxy \citep{hernquist90} is given by:
\begin{eqnarray}  
\Phi_{\star} = -\frac{GM_{\star}}{r+a},
\end{eqnarray}
where $M_{\star}$ represents the total stellar mass, and $a = R_e/1.8153$. Here, $R_e$ denotes the radius of the isophote enclosing half of the galaxy's light. The stellar mass $M_{\star}$ is determined by the virial mass $M_{\text{vir}}$ \citep{guoqi10}, as follows:
\begin{eqnarray}  
M_{\star} = 0.129M_\text{vir} C_M^{-2.44},
\end{eqnarray}
where $C_\text{M}$ is defined as:
\begin{eqnarray}  
C_\text{M} = \left(\frac{M_\text{vir}}{2.5119\times 10^{11} M_{\odot}}\right)^{-0.926} + \left(\frac{M_\text{vir}}{2.5119\times 10^{11} M_{\odot}}\right)^{0.261}.
\end{eqnarray}
The radius $R_e$ is fitted from \cite{jin20} using the relation:
\begin{eqnarray}  
R_e (\text{kpc}) = \left( \frac{0.186 M_{\star}}{10^{10} M_{\odot}} \right)^{1.116} + 1.17.
\end{eqnarray}
For this work, we have $M_{\star}  \approx 2.84\times 10^{11} M_{\odot}$ and $R_\text{e} \approx 7.59 \text{kpc}$.

The potential of the central black hole \citep{paczy80, quataert00} is given by:
\begin{equation}  
\Phi_\text{bh} = -\frac{GM_\text{bh}}{r-r_g},
\end{equation}
where $M_\text{bh}$ represents the black hole mass, and $r_\text{g} = 2GM_\text{bh}/c^2$ is the Schwarzschild radius. $M_\text{bh}$ is determined by the stellar mass  using the relation \citep{haring04}:
\begin{equation}  
M_\text{bh} = 1.6 \times 10^8 \left( \frac{M_{\star}}{10^{11} M_{\odot}} \right)^{1.12} M_{\odot}.
\label{mbh}
\end{equation}
For this work, we have $M_\text{bh} \approx 5.16 \times 10^{8} M_{\odot}$. The gravitational effect of the black hole is actually negligible in this work but not for the halo with higher mass (e.g., A1795); we include it here for consistency with future studies of larger halos.

We follow \cite{fang20} and \cite{guo20b}, with modifications, to set the distribution of the hot gas. The density profile of the gas is given by:
\begin{equation}  
\rho(r) = \frac{M_\text{n}}{(r+R_\text{e})(r+0.5R_\text{vir})^2},
\end{equation}
where $M_\text{n}$ is a constant normalized by the total gas mass $M_\text{g}$ within the virial radius, which is calculated as:
\begin{equation}  
M_\text{g} = \int_{r_\text{min}}^{r_\text{vir}} 4\pi r^2 \rho(r) \, dr.
\end{equation}
The total gas mass $M_\text{g}$ is determined by the virial mass of the dark matter halo and the gas fraction $f_\text{g}$ as $M_\text{g} = f_\text{g} M_\text{vir}$. We adopt $f_\text{g} \approx 0.108$, corresponding to a baryon fraction of about 0.1 for galaxies with a halo mass of $10^{14} M_{\odot}$, as indicated by observations \citep{vikhlinin06, lijiangtao08, zhang24}. 
When the gravitation and density profile is determined, the initial temperature profile is set up according to the assumption of hydrostatic equilibrium: $\rho \nabla \Phi = - \nabla p$ and $p =\rho k_\text{B} T/(\mu m_p)$, with the mean molecular weight $\mu = 0.61$. The temperature at the outer boundary is set as $T_\text{out} = 0.5 T_{vir} $, where the virial temperature  $T_\text{vir} $ is 
\begin{eqnarray}  
T_\text{vir} = 10^6 \left( \frac{M_\text{vir}}{ 10^{12} M_{\odot}} \right)^{2/3} K.
\end{eqnarray}
The profiles of density and temperature with different halo masses and baryon fractions are plotted in Fig. \ref{figrhoT}. We can see for the halo mass of $7.06 \times 10^{14} M_{\odot}$, this model reproduce the result fitted from observation of galaxy cluster A1795 well \citep{guo18}.   

\section{Injection of magnetic energy}
\label{app_mag}

We have developed a straightforward method to regulate the injection of magnetic energy of jets. Following \cite{mbarek19} with modifications, the toroidal magnetic field is set proportional to $r$ inside the jet base, $1/r$ outside and zero further away, as given by:

\begin{equation}  
B_{\text{inj},\ \phi} =
\left \{
\begin{aligned}
B_0 \frac{r}{R_0},\  \ & \text{for } r < R_0\ \text{and}\ z < h_{\rm jet}, \\
B_0 \frac{R_0}{r},\  \ & \text{for } R_0 < r < 2R_0\ \text{and}\ z < h_{\rm jet}, \\
0,\  \ & \text{for } r > 2R_0\ \text{or}\ z > h_{\rm jet}, \\
\end{aligned}
\right .
\label{bphi}
\end{equation}
where $R_0$ and $h_{\rm jet}$ are the cross-sectional radius and height of the jet base respectively.
Following \cite{li06} and \cite{gan17}, the poloidal magnetic field components are defined as:
\begin{equation}  
B_{\text{inj},\ z} = \alpha_p B_0 \cdot 2\left(1 - \frac{r^{2}}{R_{0}^{2}}\right) \exp\left(-\frac{r^2+z^2}{R_{0}^{2}}\right), 
\label{bz}
\end{equation}
\begin{equation}  
B_{\text{inj},\ r} = \alpha_p B_0 \cdot \frac{2zr}{R_0^2} \exp\left(-\frac{r^2+z^2}{R_{0}^{2}}\right),
\label{br}
\end{equation}
where $\alpha_p$ is the parameter of the magnetic field structure that determines the ratio between the poloidal and toroidal magnetic fields. The poloidal magnetic field is injected throughout the entire simulation region, although its value is negligible for cells far from the jet base. It is straightforward to prove that this magnetic field configuration satisfies $\nabla \cdot \textbf{B} = 0$ analytically in cylindrical coordinates. 
Then, the injection rate of magnetic energy can be expressed as:
\begin{equation}  
\dot{e}_{\text{b},\ \phi} =
\left \{
\begin{aligned}
\frac{1}{2}\dot{e}_{b0} \frac{r^2}{R_0^2},\  \ & \text{for } r < R_0\ \text{and}\ z < h_{\rm jet}, \\
\frac{1}{2}\dot{e}_{b0} \frac{R_0^2}{r^2},\  \ & \text{for } R_0 < r < 2R_0\ \text{and}\ z < h_{\rm jet}, \\
0,\  \ & \text{for } r > 2R_0\ \text{or}\ z > h_{\rm jet}, \\
\end{aligned}
\right .
\label{ebphi}
\end{equation}
\begin{equation}  
\dot{e}_{\text{b},\ z} = \frac{1}{2} \alpha_p^2 \dot{e}_{b0} \cdot 4\left(1 - \frac{r^{2}}{R_{0}^{2}}\right)^2 \exp\left(-2\frac{r^2+z^2}{R_{0}^{2}}\right),
\label{ebz}
\end{equation}
\begin{equation}  
\dot{e}_{\text{b},\ r} =  \frac{1}{2} \alpha_p^2 \dot{e}_{b0} \cdot \frac{4z^2 r^2}{R_0^4} \exp\left(-2\frac{r^2+z^2}{R_{0}^{2}}\right),
\label{ebr}
\end{equation}
where $\dot{e}_{b0} = \frac{d}{dt} (B_0^2)$. The power of magnetic energy injection is defined as:
\begin{equation}  
\begin{aligned}
P_\text{m} &=  \int_V ( \dot{e}_{\text{b},\ \phi} + \dot{e}_{\text{b},\ z} + \dot{e}_{\text{b},\ r} ) dV \\
                  &= [\ (\frac{1}{4} +\text{ln}2)\pi h_{\rm jet}R^2_0+ \frac{5\pi}{16}\sqrt{\frac{\pi}{2}} \alpha^2_p R^3_0 ]\ \dot{e}_{b0}
\end{aligned}
\label{pm1}
\end{equation}
We maintain this magnetic power as a fraction $f_\text{m}$ of the total injection power, given by:
\begin{equation}  
P_\text{m} = f_\text{m} P_\text{inj}.
\label{pm2}
\end{equation}
Combining Eqs. \ref{pm1} and \ref{pm2} the value of $\dot{e}_{b0}$ can be determined. Then, using Eqs. \ref{ebphi} - \ref{ebr}, the injection rate of magnetic energy can be solved  for all regions. The magnitude of magnetic field is updated via the equation: 
\begin{eqnarray}  
\frac{\partial B_i^2}{\partial t}  = 2\dot{e}_{\text{b},\ i},
\label{eq-B1} 
\end{eqnarray}
where $i$ represents $\phi$, $z$ or $r$. This means that the difference form of the injection rate of the magnetic field can be expressed as:
\begin{equation}  
\frac{B_i (t+\Delta t) - B_i (t) }{\Delta t} \approx \frac{ \pm [B^2_i (t) + 2\dot{e}_{\text{b},\ i}\Delta t]^{\frac{1}{2}} - B_i (t)}{\Delta t}.
\end{equation}
The sign $\pm$ is controlled to be consistent with Eq.\ref{bphi}-\ref{br}, which means $B_z$ is injected positively at $r<R_{0}$ and negatively at $r>R_{0}$, whereas $B_{\phi}$ and $B_{r}$ are both injected positively throughout the entire injection region. This method enables us to control the fraction of magnetic energy injected into the system, which is crucial for our work. To be consistent with Eq.\ref{eq-B}, Eq.\ref{eq-B1} would require: 
\begin{eqnarray}  
 \dot{B}_{\text{inj},\ i} = \frac{ \dot{e}_{\text{b},\ i}}{ B_i}. 
 \label{eq-B2} 
\end{eqnarray}

\section{Calculation of synchrotron radiation}
\label{app_rad}

Similar to some previous works, we calculate the radio emission using post-processing methods \citep{hardcastle14, yates18}. Assuming the distribution of non-thermal electrons within the radio sources follows:
\begin{equation}  
N(\gamma) = N_0 \gamma^{-q}, 
\label{N_electron}
\end{equation}
where $\gamma$ is the Lorentz factor and $q$ is the spectral index of the electrons. The spectral emission coefficient for synchrotron radiation is given by \citep{Ginzburg65, Rybicki79}:
\begin{equation}  
j_{\nu} = f(q) \frac{\sqrt{3}}{4\pi} \frac{e^3 N_0}{m_e c^2} \left(\frac{2\pi m_e c \nu}{3e}\right)^{-\frac{q-1}{2}}(B\sin \theta_p)^{\frac{q+1}{2}},
\end{equation}
where $\theta_p$ is the pitch angle of the electrons, and
\begin{equation}  
f(q) = \frac{1}{q+1} \Gamma\left(\frac{q}{4} + \frac{19}{12}\right) \Gamma\left(\frac{q}{4} - \frac{1}{12}\right).
\end{equation}

As an approximation, the distributions of electrons and radiation are typically assumed to be locally isotropic, allowing the emission coefficient to be expressed as:
\begin{equation}  
j_{\nu} = a(q) \frac{\sqrt{3}}{4\pi} \frac{e^3 N_{0}}{m_e c^2} \left(\frac{2\pi m_e c \nu}{3e}\right)^{-\frac{q-1}{2}} B^{\frac{q+1}{2}}
\end{equation}
where,
\begin{equation}  
a(q) = \frac{\sqrt{\pi}}{2} \Gamma\left(\frac{q+5}{4}\right) \left[\Gamma\left(\frac{q+7}{4}\right)\right]^{-1} f(q)
\end{equation}

\begin{figure*}
\centering
\includegraphics[height=0.24\textheight]{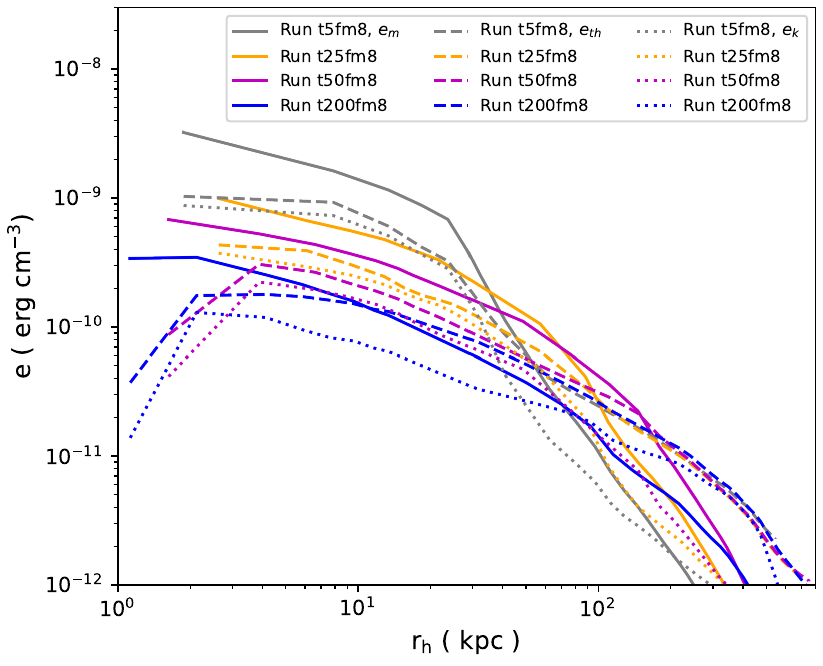}
\includegraphics[height=0.24\textheight]{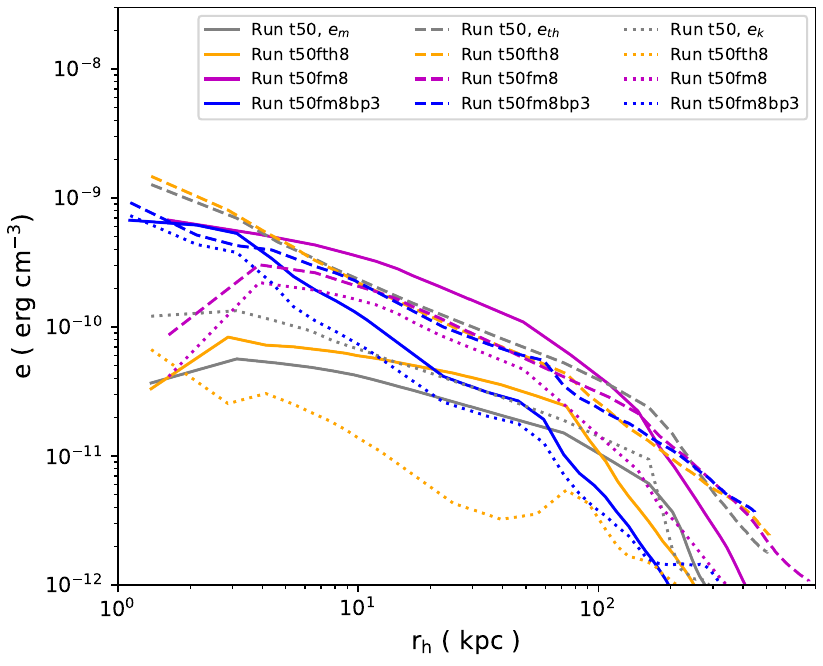}
\caption{Volume-averaged energy density of different energy components varying with the traveling distance of lobes in the two sets of runs discussed in Sect. \ref{sec4.2}. Different colors represent different runs, whereas the dotted, dashed, and solid lines signify kinetic, thermal, and magnetic energy densities, respectively.}
\label{figei}
\end{figure*}

The radiation exhibits a spectral distribution of $j_{\nu} \sim \nu^{-\alpha}$ with $\alpha = (q-1)/2$. We adopt $\alpha = 0.7$, a typical value in the observation \citep{dabhade23}. Given that the magnetic field can be derived from our simulations, the only unknown parameter is $N_0$ in Eq. \ref{N_electron}. According to Eq. \ref{N_electron}, the energy density of non-thermal electrons is given by:
\begin{equation}  
u_e = \int_{\gamma_{\text{min}}}^{\gamma_{\text{max}}} N(\gamma) \gamma m_e c^2 \, d\gamma.  
\end{equation}
Thus, $N_0$ can be determined as follows:
\begin{eqnarray}  
N_{0} = \left \{
\begin{aligned}
&\frac{u_e}{m_ec^2} \ln( \frac{\gamma_\text{min} }{ \gamma_\text{max} })&,    \ \  q=2 \\ \\
&\frac{u_e}{m_ec^2}  \frac{2-q}{ \gamma_{max}^{(2-q) }- \gamma_{min}^{(2-q)}}&,    \ \  q\neq 2 \\
\end{aligned}
\right .
\label{N_0}
\end{eqnarray}
Here, $\gamma_{\text{min}}$ and $\gamma_{\text{max}}$ are empirically set to $100$ and $10^{5}$, respectively \citep{hardcastle14, hardcastle18}. We assume that $u_e$ is a fraction $\eta_e$ of the non-magnetized energy, defined as:
\begin{equation}  
u_e = \eta_e \left(e - \frac{B^2}{2}\right),   
\label{u_e}
\end{equation}
where $e$ represents the total energy density in Eq. \ref{eq-e}. For self-consistency, we adopt $\eta_e = 0.1$ as we expect the non-thermal electrons to have negligible impact on the dynamics of the jets and lobes in our simulations. Finally, $N_0$ can be derived from Eqs. \ref{N_0} and \ref{u_e}. In Fig. \ref{figei}, we show the volume-averaged energy density  of various energy components varying with the traveling distance of the lobes. 

\begin{figure*}
\centering
\includegraphics[height=0.28\textheight]{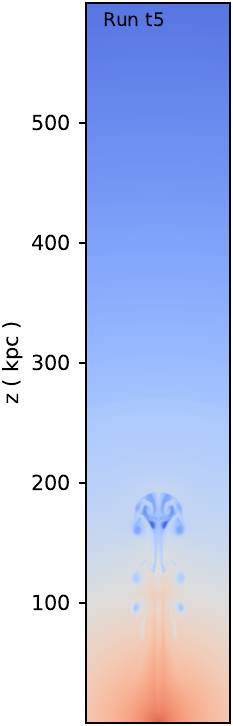}
\includegraphics[height=0.28\textheight]{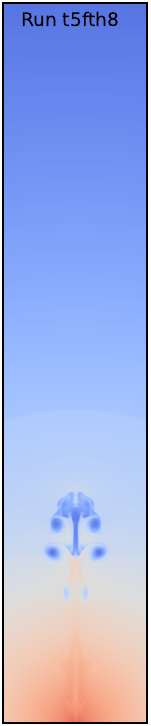}
\includegraphics[height=0.28\textheight]{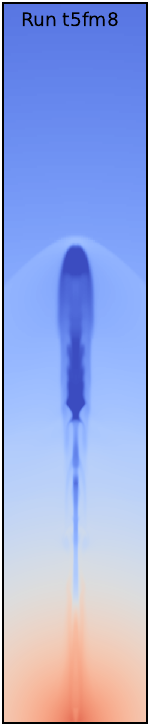}
\includegraphics[height=0.28\textheight]{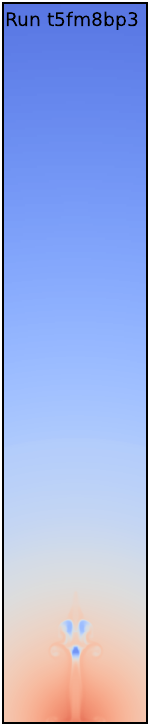}
\includegraphics[height=0.28\textheight]{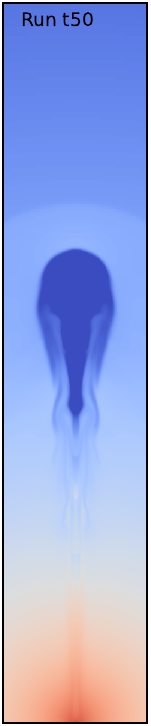}
\includegraphics[height=0.28\textheight]{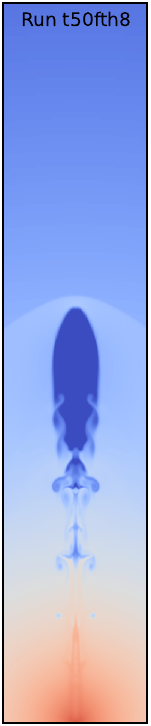}
\includegraphics[height=0.28\textheight]{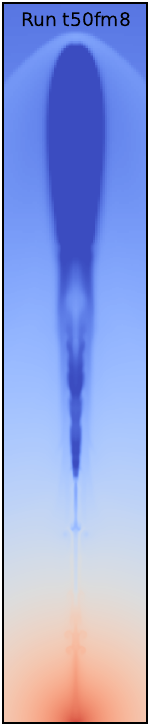}
\includegraphics[height=0.28\textheight]{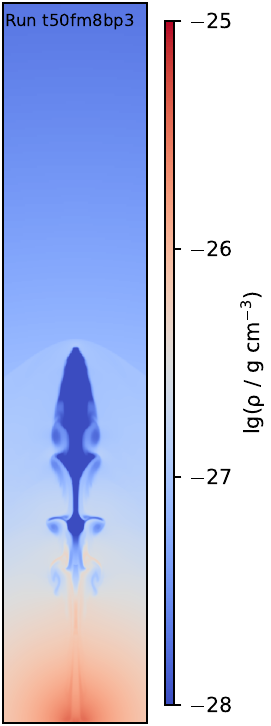}\\
\includegraphics[height=0.282\textheight]{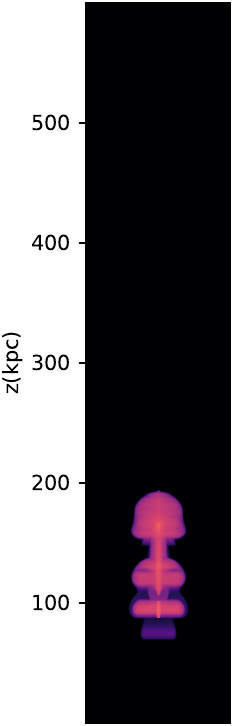}
\includegraphics[height=0.282\textheight]{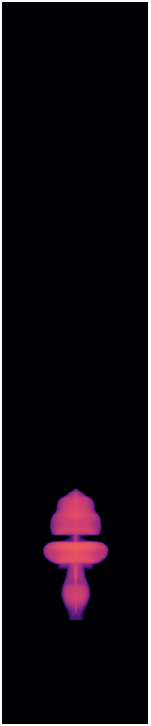}
\includegraphics[height=0.282\textheight]{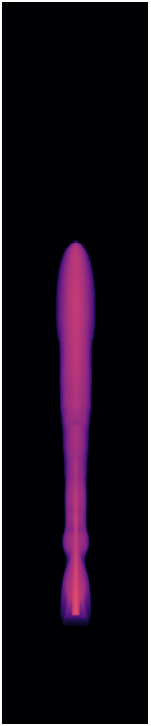}
\includegraphics[height=0.282\textheight]{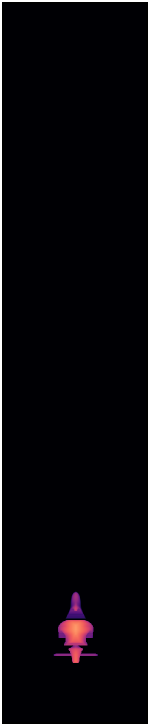}
\includegraphics[height=0.282\textheight]{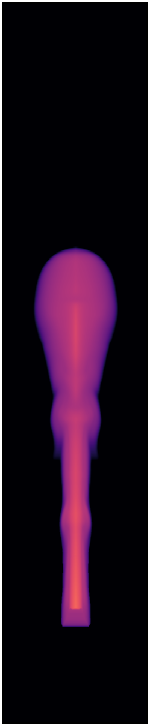}
\includegraphics[height=0.282\textheight]{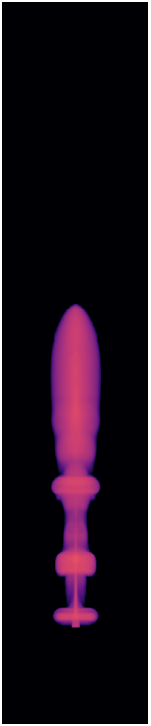}
\includegraphics[height=0.282\textheight]{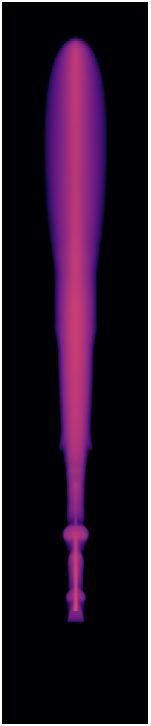}
\includegraphics[height=0.282\textheight]{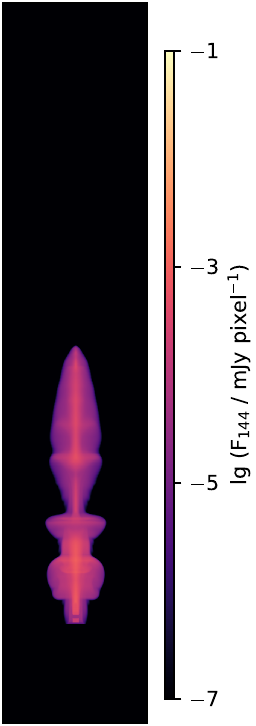}\\
\includegraphics[height=0.282\textheight]{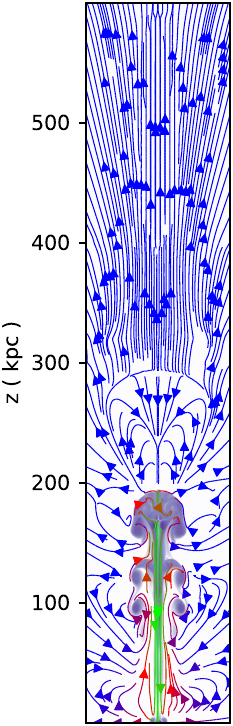}
\includegraphics[height=0.282\textheight]{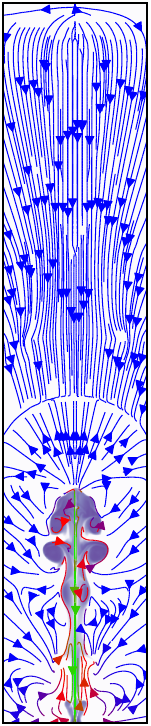}
\includegraphics[height=0.282\textheight]{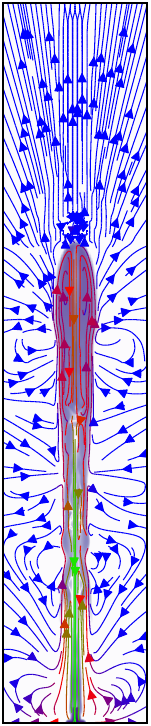}
\includegraphics[height=0.282\textheight]{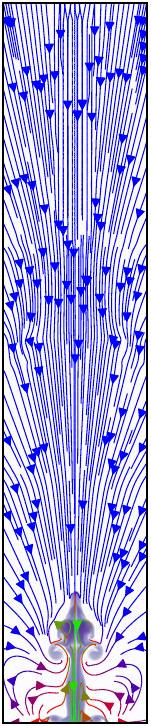}
\includegraphics[height=0.282\textheight]{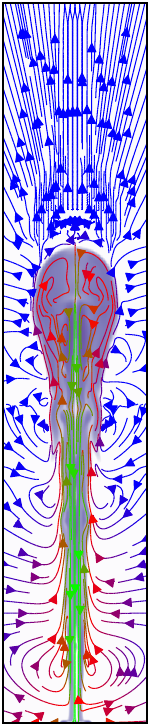}
\includegraphics[height=0.282\textheight]{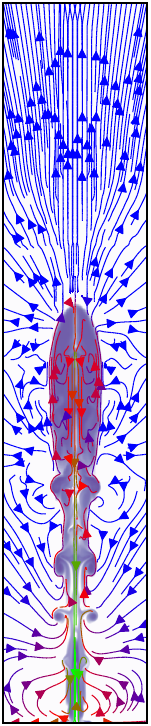}
\includegraphics[height=0.282\textheight]{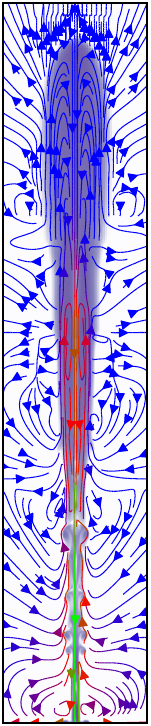}
\includegraphics[height=0.282\textheight]{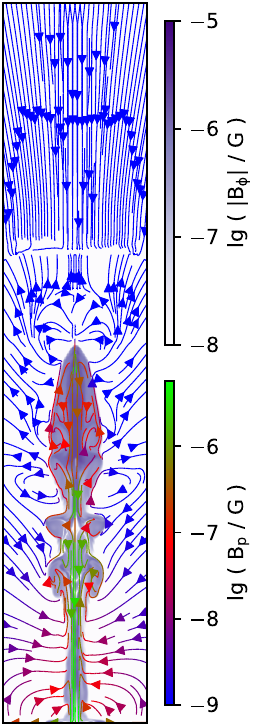}
\caption{Snapshots of density (top), radio flux (middle), and magnetic field (bottom) for runs with $t_{\text{jet}} = 5 \text{Myr}$ and $t_{\text{jet}} = 50 \text{Myr}$ at $t = 300 \text{Myr}$. The pixel sizes in the radio flux plots directly mirror the cell sizes employed in the simulations.}
\label{fig_morph2}
\end{figure*}

The radio power (luminosity) of radio sources can be calculated as:
\begin{equation}  
P_{\nu} = \int_V 4\pi j_{\nu} \, dV.
\end{equation}
The radio flux can be determined by:
\begin{equation}  
\Delta F = \int_{l_-}^{l_+} j_{\nu} \, dl \cdot \frac{\Delta A}{d_{L}^2},
\end{equation}
where $l_- - l_+$ represents the integral path traversing the radio sources along the line of sight perpendicular to the jet axis. $\Delta A$ denotes the cross-sectional area of the simulation cells along the line of sight, and $d_L$ is the luminosity distance. For this study, we place the sources in our simulations at a redshift of $z=0.3$ (typical value of the sources in \cite{dabhade20a}) and calculate the radio power and flux at frequency $\nu = 144 \rm (1+z) MHz$. The radiating lobe region is identified by a density lower than $5\times 10^{-27} \rm{g\ cm^{-3}}$ and a toroidal magnetic field stronger than $10^{-9} \rm{G}$. 
In Figs. \ref{fig_morph1} and \ref{fig_morph2}, we present snapshots of the radio flux distribution of the sources from selected runs, alongside the corresponding distributions of density and magnetic field. 

The synchrotron cooling time of electrons in the radio lobes of AGN is typically hundreds of million years \citep{Ghisellini13}. The cooling time at a certain frequency \citep{Courvoisier13} can be estimated as 
\begin{equation}  
t_{\rm syn}(s) \approx 6\times 10^8 B^{-3/2}_{\rm G} \nu^{-1/2}_{\rm MHz}.
\label{scooling}
\end{equation}
For the frequency of $\rm \nu=144(1+0.3)MHz$, Eq. \ref{scooling} results in a cooling time of about 43.9 Myr for a magnetic field of $10^{-5}$ G and 1.4 Gyr for $10^{-6}$ G, which means the electrons will undergo significant synchrotron cooling if the magnetic field is sufficiently strong when the jets are shut off.

\section{Integral energy equation in Lagrangian form}
\label{app_etran}

Here we consider the energy loss from a given jetted bubble as it interacts with the surrounding medium. The energy equation Eq.\ref{eq-e} can be written in its full conservation form, as usually shown in textbooks \citep{Ogilvie16}. Omitting the jet injection, we have:
\begin{equation}  
\frac{\partial }{\partial t}(e+\rho \Phi) + \nabla \cdot [ (e + \rho \Phi) \textbf{v} +(p + \frac{B^2}{2})\textbf{v}- \textbf{B} \textbf{B} \cdot \textbf{v} ] = 0 .
\label{eq-e2}
\end{equation}  
When considering the energy outflow from a moving and growing bubble, we must refer to the integral energy equation in Lagrangian form. While neglecting the mixing processes, the energy outflow rate from the region $V_{bub}$ of the given bubble can be written as 
\begin{equation}  
\begin{aligned}
  P_{out} &= - \frac{d}{dt} \int_{V_{bub}} (e+\rho \Phi) dV \\
               &= - \int_{V_{bub}} [ \frac{d }{d t}(e+\rho \Phi) + (e+\rho \Phi) \nabla \cdot \textbf{v}] dV \\
               &= - \int_{V_{bub}} \{ \frac{\partial }{\partial t}(e+\rho \Phi) + \nabla \cdot [(e+\rho \Phi) \textbf{v}] \} dV .
\end{aligned}
\label{eq-pout1}
\end{equation}  
Here we have used the relation between the rate of change of a small volume and velocity $\frac{1}{\Delta V}\frac{d\Delta V}{dt} = \nabla \cdot \textbf{v} $ and the relation between Lagrangian and Eulerian derivatives $\frac{d}{dt} = \frac{\partial}{\partial t} + \textbf{v}\cdot \nabla$ (or applying the Reynolds Transport Theorem). Then, using Eq.\ref{eq-e2}, we have
\begin{equation}  
 P_{out} = \int_{S_{bub}} [(p + \frac{B^2}{2})\textbf{v}- \textbf{v} \cdot \textbf{B} \textbf{B}  ] \cdot d\textbf{S}, 
\label{eq-pout2}
\end{equation}  
where $S_{bub}$ is the surface of the bubble. The physical meaning is obvious: the bubble mainly loses energy through adiabatic processes due to the work done by the thermal pressure and Maxwell stress tensor (magnetic pressure plus magnetic tension). The same derivation for the mass equation results in a zero mass loss rate, indicating the mass conservation in the moving bubbles when the mixing processes are neglected.

\end{appendix}

\end{document}